\documentclass[final,5p,times,twocolumn]{elsarticle}
\usepackage{amssymb}
\usepackage{graphicx}
\usepackage[figuresright]{rotating}
\usepackage{color}

\journal{Physics Letters B}
\begin{document}
\begin{frontmatter}

\title{Evolution of fusion hindrance for asymmetric systems at deep  sub barrier energies }

\author[barc,hbni]{A.~Shrivastava \corref{cor1}}
\cortext[cor1]{Corresponding author}
\ead{aradhana@barc.gov.in}
\address[barc]{Nuclear Physics Division, Bhabha Atomic Research Centre, Mumbai 400085, India}
\address[hbni] {Homi Bhabha National Institute, Anushaktinagar, Mumbai 400094, INDIA}
\author[barc,hbni]{K.~Mahata}
\author[barc,hbni]{S.K.~Pandit}
\author[TIFR]{V.~Nanal}
\address[TIFR] {DNAP, Tata Institute of Fundamental Research, Mumbai 400005, India}
\author[kyoto]{T.~Ichikawa}
\address[kyoto]{Yukawa Institute for Theoretical Physics, Kyoto University, Kyoto 606-8502, Japan}
\author[Tohuku]{K.~Hagino}
\address[Tohuku]{Department of Physics, Tohuku University, Sendai 980-8578, Japan}
\author[GANIL] {A. Navin}
\address[GANIL] {GANIL, CEA/DRF - CNRS/IN2P3, Bd Henri Becquerel, BP 55027, F-14076 Caen  Cedex 5, France} 
\author[barc]{C.S.~Palshetkar}
\author[barc]{V.V.~Parkar}
\author[barc,hbni]{K.~Ramachandran}
\author[barc,hbni]{P.C.~Rout}
\author[barc]{Abhinav Kumar}
\author[barc]{A.~Chatterjee}
\author[barc]{S.~Kailas}

\begin{abstract}

Measurements  of fusion cross-sections of $^{7}$Li and $^{12}$C with $^{198}$Pt  at deep sub-barrier energies are reported to unravel the role of the entrance channel in the occurrence of fusion hindrance. The onset of fusion hindrance has been clearly observed in $^{12}$C~+~$^{198}$Pt system but not in $^{7}$Li~+~$^{198}$Pt system,  within the measured  energy range.   Emergence of the  hindrance,  moving from lighter ($^{6,7}$Li) to heavier ($^{12}$C,$^{16}$O) projectiles is  explained  employing a   model that considers a gradual transition from a sudden to adiabatic regime at low energies.  The  model calculation reveals a weak effect of the damping of coupling to collective motion for the present systems  as compared to that obtained for systems with heavier projectiles.
\end{abstract}
 

\end{frontmatter}

\section{Introduction} 
Fusion reactions in the vicinity of the Coulomb barrier have been investigated in the past to explore  the mechanism of tunneling through  multidimensional barriers, thereby giving an insight into    the role of  different intrinsic properties of the entrance channel.  Recent efforts towards developing new methods to precisely measure very low fusion cross-sections have stimulated new activities, distinct to energies deep below the barrier. Fusion data at these low energies can be uniquely  used to interpret the reaction dynamics from the touching point to the region of complete overlap of the density distribution of the colliding nuclei, not accessible through any other reaction~\cite{review,review2}.  This opens up the possibility to study   effects of dissipative quantum tunneling, which has relevance in many fields of physics and chemistry~\cite{dissi-prl}. The data in this energy range was   shown to have  strong implications on the fusion  with light nuclei of astrophysical  interest~\cite{review2}.

At deep sub-barrier energies, a change of slope of  the fusion excitation function compared  to coupled-channels (CC) calculations was observed initially  in symmetric systems involving medium-heavy nuclei and was referred to as the phenomenon of fusion hindrance~\cite{jia02,jia04}. The models suggested to explain this behavior have different physical basis. The  model proposed by Mi\c{s}icu and Esbensen is  based  on a  sudden approximation~\cite{esb03}, where a repulsive core is included to take into account the nuclear compressibility arising  due to Pauli exclusion principle when the two nuclei overlap.    On the other hand  at low energies, the  nucleus-nucleus  interaction potentials extracted from  the microscopic time-dependent Hartree-Fock theory indicate that after overlap of two nuclei, internal degrees of freedom  reorganise adiabatically~\cite{washiyama}. The model proposed  by Ichikawa  {\it et al.}~\cite{hag07}  to explain the deep sub-barrier  fusion data is based on such an adiabatic picture. Here  a   damping factor imposed  on the coupling strength  as a function of the inter-nuclear distance, takes into account a gradual change from the sudden to the adiabatic  formalism~\cite{ichikawa2009, ichikawa_sub}. A recent work, applying the random-phase-approximation (RPA) demonstrates that the fusion hindrance originates from damping of quantum vibrations when the two nuclei adiabatically approach each other~\cite{ichikawa2013,ichikawa2015}.  The role of quantum de-coherence  that effectively cause a reduction in coupling effects has also been investigated ~\cite{Daiz08,maha07}.

In all the above models, fusion hindrance is a generic property of heavy-ion collision below certain threshold energy. Due to challenges involved with measurement of low cross-section ($\sim$nb),   there are only a limited number of studies involving   fusion hindrance.  As discussed in a recent review article~\cite{review2}, these studies   have mainly concentrated around medium-heavy (A$\sim$100), medium (A$\sim$50) and   light (A$\sim$10)  symmetric systems~\cite{jia02,jia04,jia05,stef10,jia10,mont2010,mont2013,stef2014,jiang14}, covering  a  wide  range   of  reduced   masses, Q-values and nuclear structure properties.   Most of the measurements employed  recoil  mass analyzers and hence are restricted to symmetric or nearly symmetric systems.  In such cases the evaporation residues have sufficient recoil velocities for being detected at the focal plane of the spectrometer. The data corresponding to asymmetric systems, presently scarce,   are vital to establish the generic nature of the fusion hindrance  and for the improvement of current theoretical models. The only exception being the two   systems $^{16}$O~+~$^{208}$Pb~\cite{maha07} and $^6$Li~+~$^{198}$Pt~\cite{araPRL} that used different methods for fusion cross-section measurements. The presence of fusion hindrance was clearly shown in $^{16}$O~+~$^{208}$Pb system~\cite{maha07}. The shapes of the logarithmic derivative and astrophysical S-factor for this asymmetric system were found to be different, compared to those for the symmetric systems~\cite{review,review2}. In the case of a more asymmetric system  $^6$Li~+~$^{198}$Pt~\cite{araPRL}, an absence of fusion hindrance was reported at energies  well below the threshold energy (E$_T$) computed from both  the sudden and adiabatic models.  For reactions induced by protons, intuitively one would not expect fusion hindrance.  In this case, the projectile maintains its identity and the sudden approximation would be appropriate. This should be the case for alpha particle as well, which can be treated as a rigid nucleus. On the other hand for heavier projectiles, such as $^{12}$C and $^{16}$O, one may expect a neck formation  at low energies when the colliding nuclei follow the minimum energy path allowing for the readjustment of the densities as a function of the collective variables. Deviation from a simple sudden picture is expected to occur for nuclei heavier than $^4$He.

\begin{figure}
\includegraphics[width=17pc]{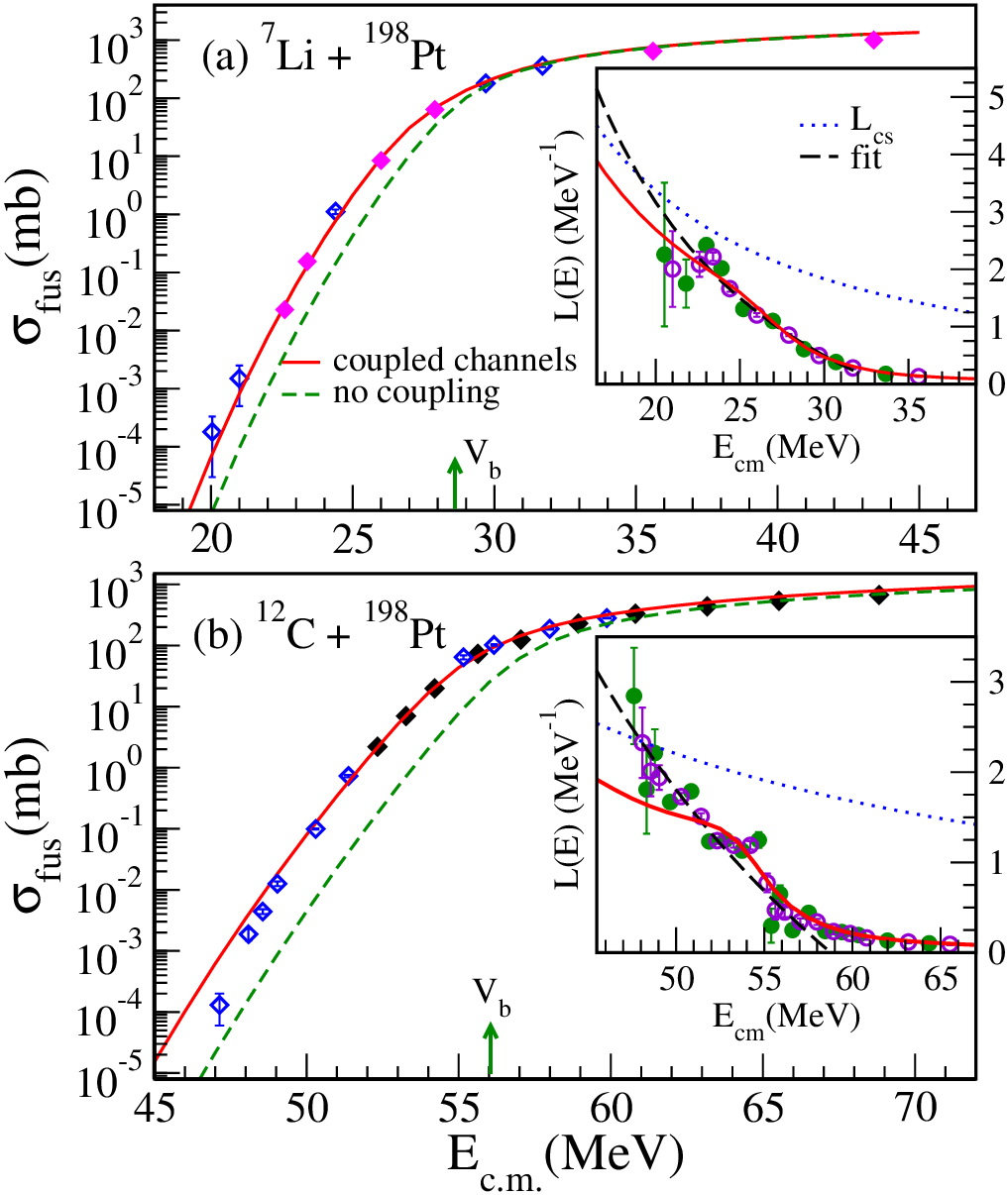}
\caption{  (Color  online) Fusion  excitation  function and its logarithmic derivative (inset)  for  (a) $^{7}$Li~+~$^{198}$Pt and  (b) $^{12}$C~+~$^{198}$Pt  systems. The arrow indicates the value of the  Coulomb barrier (V$_b$). The cross-sections  from Refs.~\cite{araPRC,araPLB} are shown as filled diamonds. The L(E) values shown as closed and open circles were obtained from two consecutive data points  and   least-squares fits to three successive data points, respectively.   The  results  of the  coupled-channels  calculations  (solid  line) along  with single   channel calculations (dashed line)  using the code {\sc CCFULL}   are  also shown. The L(E) values fitted to an expression and that corresponding to a constant S-factor (L$_{cs}$(E)) are shown as long dashed and dotted curves, respectively (see text).  }
\label{fig:fig1}
\vskip -5mm
\end{figure}

  The present work investigates the evolution of the fusion hindrance with increasing mass and charge of relatively light projectiles ($^{6,7}$Li, $^{12}$C, $^{16}$O) on heavy targets. For this purpose we have performed new measurements at deep sub-barrier energies with $^7$Li and $^{12}$C projectiles  on a $^{198}$Pt target.  The current results along with the available data for different entrance channels have been studied to understand the origin of the fusion hindrance. 

\section{Experimental details and results}
 The  experiments were performed  at the Pelletron-Linac  Facility, Mumbai, using beams of $^{7}$Li (20 - 35 MeV)  and $^{12}$C (50 - 64 MeV)  on a $^{198}$Pt target with beam current in the range of 10 to 35~pnA.   The targets were  foils of $^{198}$Pt (95.7$\%$ enriched, $\sim$ 1.3~mg/cm$^2$ thick) followed by an Al  catcher foil of  thickness $\sim$ 1~mg/cm$^2$.   The  cross-sections have been extracted using a sensitive and selective offline method employing KX-$\gamma$ ray coincidence~\cite{araPRL,ant09}. Two efficiency calibrated HPGe detectors - one with an Al window for detection of $\gamma$-rays and another  with a Be window for detection of KX-rays,  having an active volume $\sim$ 180cc were placed  face to face for performing KX-$\gamma$-ray coincidence of  the decay radiations from the  irradiated sample.  The irradiated targets  were mounted at $\sim$ 1.5~mm from the face of each detector. The  measurements were  performed in  a  low background  setup with  a graded shielding (Cu,  Cd sheets of thickness $\sim$  2~mm followed by 10~cm Pb).   The evaporation residues  from complete fusion were uniquely identified by means of their characteristic $\gamma$-ray energies and half-lives which correspond to   $^{205-207}$Po  in case of $^{12}$C~+~$^{198}$Pt and $^{200-202}$Tl in case of $^7$Li~+~$^{198}$Pt systems. The  $\gamma$-ray yields of the daughter nuclei at lowest energies were extracted by gating on their KX-ray transitions.   Further   details  on  the  method  can   be  found  in Ref. \cite{ant09}.   Due to the increased sensitivity of the  KX-$\gamma$-ray coincidence  method,  cross-section  down  to  130 nano-barns  could  be  measured. The  fusion cross-sections  were obtained  from the  sum  of the  measured evaporation residue  cross-sections. In case of $^{12}$C~+~$^{198}$Pt system, the fission cross-section was also taken into account  using data from Ref.~\cite{araPRC}, up to the beam energy where fission cross-section was $\sim$ 0.5 \% of the fusion cross-section. The statistical  model calculations for the compound  nuclear decay were performed using PACE \cite{gav80} with parameters  from Refs.~\cite{araPRL,araPRC}  which   reproduce  the residue cross-sections   well for both the systems.  The  estimation of  errors  for  low counting  rates was  made assuming  Poisson statistics  and  using the method  of  maximum likelihood~\cite{yao06}.  The present results are shown in Fig.~\ref{fig:fig1}, together with the cross-sections obtained in Refs.~\cite{araPRC,araPLB}. The error on the data points in Fig.~1 is only  statistical in nature. 

   Plotted  in the inset of Fig.1(a) and (b)  are  the  logarithmic derivatives  of  the  fusion cross-section  (L(E)=d[ln($\sigma$E)]/dE), determined using two consecutive data points and also performing a least square fit to a set of three data points. This  representation provides an alternate  way to illustrate any  deviations in the  slope of the  fusion excitation function independent of the weight of  the lowest barrier~\cite{review2}.  The L(E) values fitted to the expression (A~+~B/E$^{3/2}$)  and that corresponding to a constant astrophysical S-factor (L$_{cs}$(E))~\cite{jia06} are shown as long dashed and dotted lines respectively. The cross-over point between the L(E) and L$_{cs}$(E) corresponds to peak of the S-factor and can be related to the threshold  energy for observing fusion hindrance~\cite{jia06}.

\section{Calculations}

  Coupled-channels  calculations   using  the   code   {\sc CCFULL}~\cite{hagino}   were    performed   for both the systems. In the case of $^7$Li~+~$^{198}$Pt system, a standard Woods-Saxon potential  (WS) was used with V$_0$=110~MeV,  r$_0$=1.1~fm  and $a$=0.63~fm.    These  calculations included   two phonon quadrupole excitation of $^{198}$Pt in the vibrational  and the first excited state  of $^{7}$Li  in the rotational mode.    The  CC calculations reproduce the data well for energies around and well below the barrier as seen in Fig.~1(a). Fusion hindrance has not been observed in this system in the measured energy range with the  cross-section as low as  $\approx$ 180 nb.   The threshold energy   for observing fusion hindrance obtained from the systematics of Ref.~\cite{jia09}  and the adiabatic model~\cite{hag07} is 20.4~MeV and 21.1~MeV, respectively.    However, from an extrapolation of the experimental data, this energy  is found to be $\approx$19~MeV (Fig.~1(a) inset). Hence it will be interesting to extend the measurement of fusion cross-sections below the lowest energy of the present measurement (20 MeV). 

  \begin{figure}

\includegraphics[width=21pc]{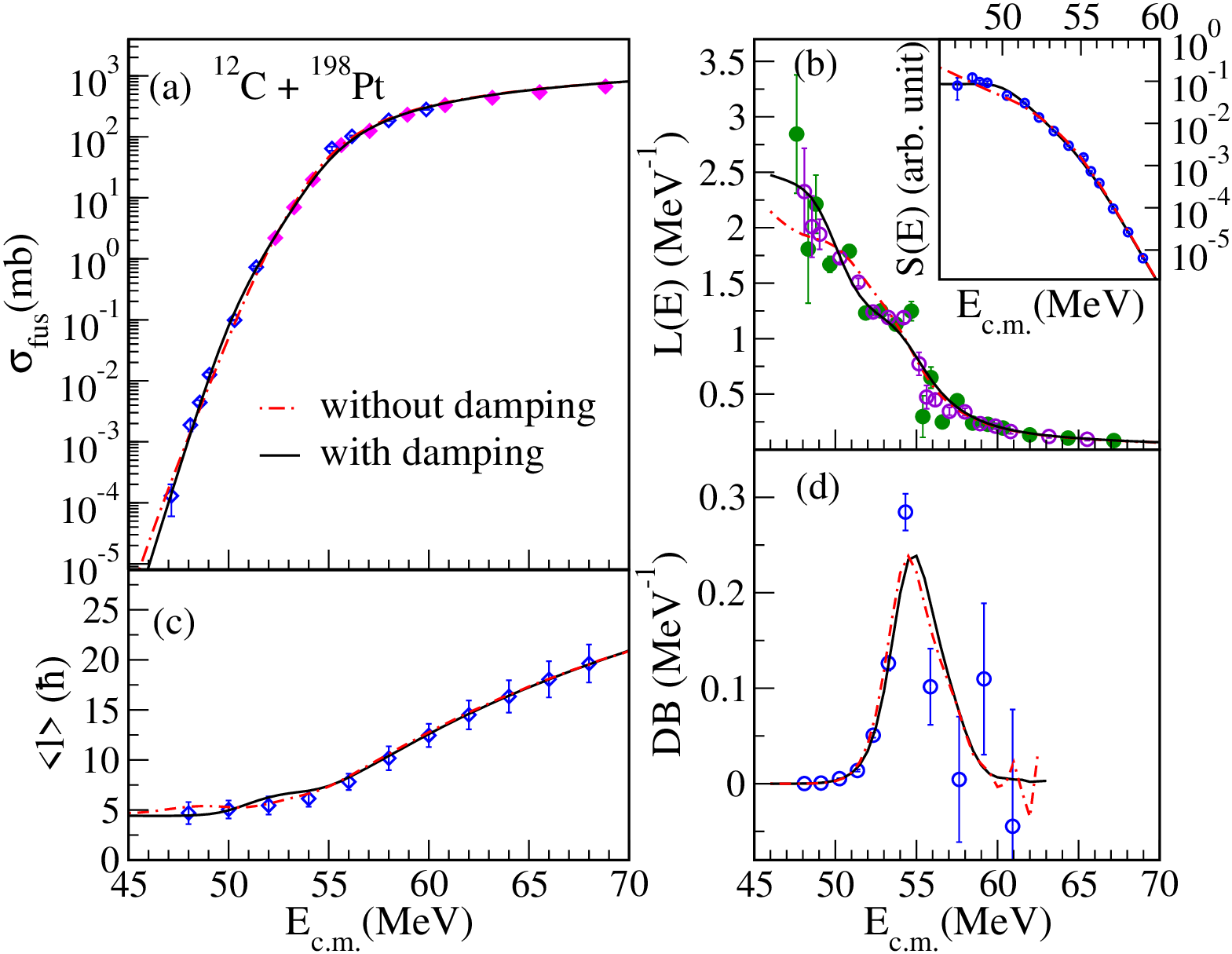}
\caption{  (Color  online) Results from the adiabatic model calculation  for $^{12}$C~+~$^{198}$Pt system compared with the experimental (a)  fusion cross-sections,   (b)    logarithmic derivative along with  S-factor  (inset) (c) average angular momentum and (d) fusion barrier distribution. Calculations using  with and without a damping factor for the coupling strength  are shown as solid and dashed-dot curves respectively.}

\label{fig:fig2}
\vskip -5mm
\end{figure}

  The corresponding calculations  for $^{12}$C~+~$^{198}$Pt were performed  using a WS potential with V$_0$=95~MeV,  r$_0$=1.13~fm  and $a$=0.66~fm. The coupling to the  quadrupole  phonon excitation for  $^{198}$Pt and the first two excited states of $^{12}$C belonging to the ground state rotational band were included. The     quadrupole and hexadecapole deformation parameters used were taken from   Ref.~\cite{yas83}. The effect  of   coupling to the $^{12}$C rotational  states is not as strong  as in the well deformed heavy nuclei. Coupling to one neutron, two neutron and one proton transfer reaction were not included in the present scheme as their effect was found to be negligible   for this system~\cite{araPRC}. The result from the CC calculations are compared with the experimental data in Fig.~1(b).  A change of slope as compared to CC calculations is clearly observed,  both in the measured fusion excitation function as well as in the L(E) plot, confirming the onset of fusion hindrance. The energy at which the deviation in the slope occurs was estimated to be 50~$\pm$~1 MeV using the method described in Ref.~\cite{Ei12}.  The calculated threshold energy   according to the  adiabatic model~\cite{hag07}  is 49 MeV  while that from the systematics (43.7 MeV)~\cite{jia09} is much lower than the observed value.

  In order to explain the fusion data at energies deep below the barrier in case of $^{12}$C~+~$^{198}$Pt system,  calculations were performed using  the adiabatic model of Ref.~\cite{ichikawa2009,ichikawa_sub}. This model employs a gradual diminishing of the coupling strength  while going from the two body sudden to one body adiabatic potential as the two nuclei begin to overlap. The calculations adopted a Yukawa-plus-exponential (YPE) potential  as a basic ion-ion potential with  radius, r$_0$=1.20 fm and  diffuseness, $a$=0.68 fm. The coupling scheme was the same as that described earlier for this system. The calculated fusion cross-sections without damping, shown as the dot-dashed curves in Fig. 2(a), already provide a good fit, although the calculation underestimates the data for L(E) at the lowest energies (see Fig. 2(b)).  The calculations shown here differ slightly from those in Fig~1(b). This is due to the use of different potentials (YPE and Woods-Saxon) in these two calculations, and the fact that the YPE potential is thicker  than the Woods-Saxon potential (due to the saturation condition at the touching point in the YPE potential). Further discussion about  the choice of potentials used in the present work can be found in section IIC of Ref.~\cite{ichikawa_sub}.  Fig.~2 also shows the results of the calculation with the inclusion of a damping factor (r$_{damp}$=1.18 fm and $a_{damp}$=0.5~fm),  which are in excellent agreement with both the fusion and the L(E) data. As can be seen from the figure, the effect of the damping  is observed to be small in the present case when compared to that observed in studies involving  heavier projectiles~\cite{ichikawa_sub}.  A systematic investigation of  various systems showed that the radius parameter, related to the density distribution of the colliding nuclei, is  almost constant~\cite{ichikawa_sub}. On the other hand $a_{damp}$, associated with the damping strength of quantum vibrations that depends on the structure of interacting nuclei,  was found to vary between 0.5 and 1.2 fm. The values of r$_{damp}$ and $a_{damp}$ obtained in the present work are within the range of the values reported in Ref.~\cite{ichikawa_sub}. 

 The adiabatic calculations were compared with other observables derived from the fusion data.  The astrophysical S-factor representation (S(E)) of the experimental data is shown in the inset of Fig.~2(b).  The observed S-factor maximum  is not as pronounced  as found for the case of the  symmetric systems involving medium mass nuclei, but similar to that for   $^{16}$O~+~$^{208}$Pb system~\cite{review,review2,Jia07}.  The  calculated S(E)   match well with the data over the entire energy range. The average  angular momenta  ($\langle l \rangle$)  computed from    the    fusion   excitation    function    as   suggested    in Ref.~\cite{bala} and the fusion barrier distribution (DB) are also well described by the adiabatic calculation (Fig.~2(c) and (d)). 

 Similar calculations were performed for $^{7}$Li~+~$^{198}$Pt  using the YPE potential (r$_0$=1.195 fm and $a$=0.68 fm) with the coupling scheme being the same as that described above for this system. The calculations explain the data well up to the measured energy. The values of r$_{damp}$ and $a_{damp}$ can not  be  determined uniquely with the present data as no change of slope of the fusion excitation function was observed. For example,  values of the damping factor parameters r$_{damp}$=1.16 fm and $a_{damp}$=0.5 fm,  would give rise to a small deviation in the slope at energy $\sim$ 19 MeV. The threshold energy for observing hindrance is expected to be below this value. 

\section{Discussion}
    We now discuss the general trend of fusion excitation function at deep sub-barrier energies for asymmetric systems involving light projectiles, namely,  $^{6}$Li~+~$^{198}$Pt~\cite{araPRL},  $^{7}$Li~+~$^{198}$Pt, $^{12}$C~+~$^{198}$Pt and $^{16}$O~+~$^{208}$Pb  ~\cite{maha07,morton}. $^{6,7}$Li+$^{198}$Pt are among the few systems,  that have been probed for hindrance studies, having positive Q-values for the formation of compound nucleus~\cite{review2}. As $^{6,7}$Li are weakly bound nuclei  ($^6$Li, S$_{\alpha/d}$=1.47 MeV and $^7$Li, S$_{\alpha/t}$=2.47 MeV), the role of the breakup channel at energies relevant to the fusion hindrance needs to be considered as well. The influence of breakup on fusion and total reaction cross-sections has been extensively investigated~\cite{Gomes,yang}. Recent studies have also illustrated  the importance of   transfer followed by breakup channels~\cite{luong,ara06}. However, inclusion of such processes simultaneously in a coupled channels framework to predict complete fusion cross-section is still a challenging task~\cite{Gomes}.

To study the onset  of the fusion hindrance for asymmetric systems involving light projectiles, the ratio~\cite{jia02} of experimental fusion cross-section to that obtained from the standard CC calculations are shown in Fig.~3(a).   The ratio  remains close to one at near and deep sub-barrier energies in case of systems involving the lightest projectiles $^{6,7}$Li, showing no deviation even at energy as low as  $\sim$ 10 MeV below the barrier. However, for the heavier projectiles $^{12}$C and $^{16}$O, there is a significant change in the slope with respect to the calculations  at the lowest energies (V$_B$~-~E$_T$$\sim$ 6 MeV). The fusion hindrance becomes gradually larger in moving from lighter ($^{6,7}$Li) to the relatively heavier projectiles ($^{12}$C and $^{16}$O).

\begin{figure}
\includegraphics[width=21pc]{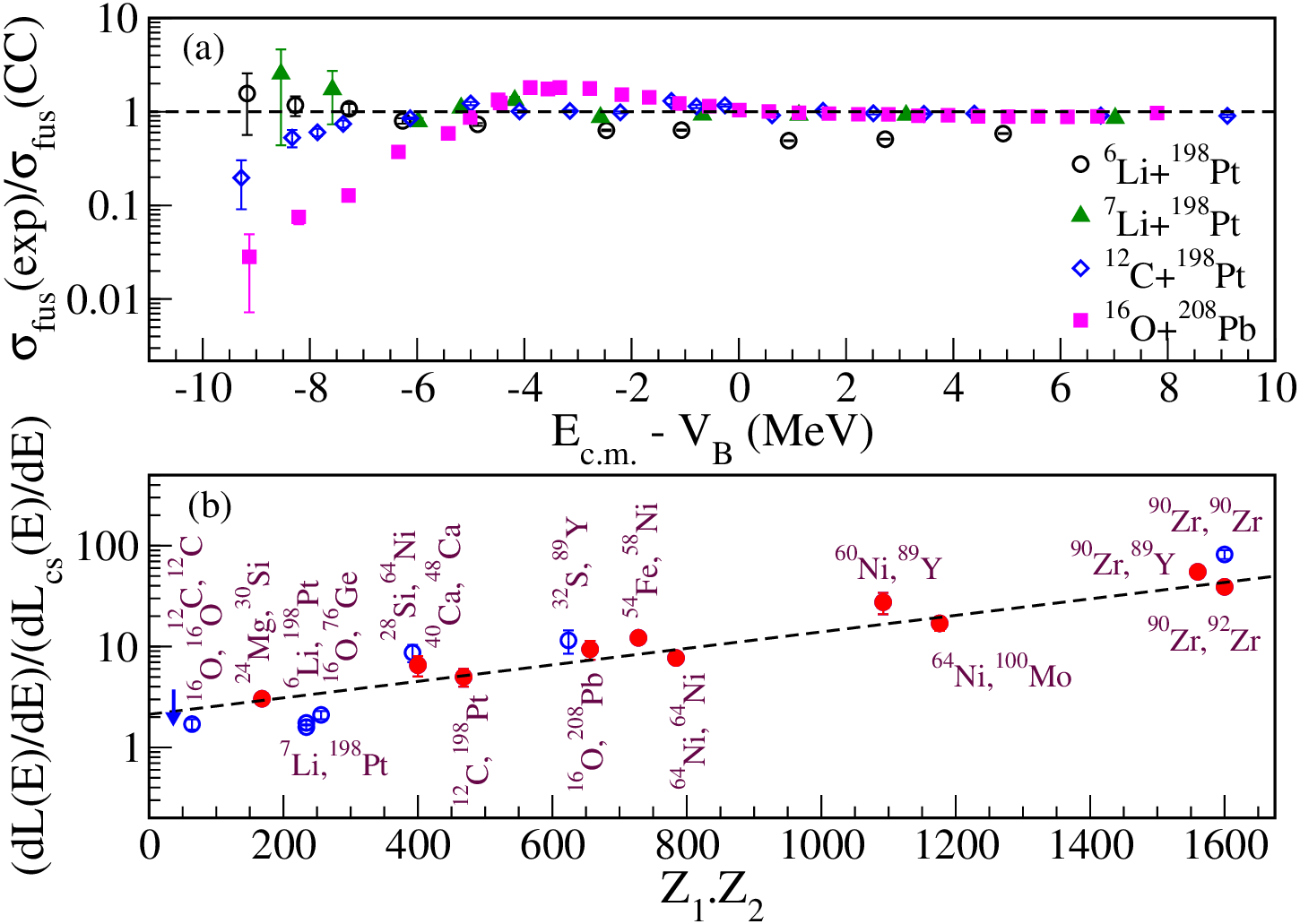}
\caption{  (color  online)  (a) Ratio of the measured and calculated fusion cross-sections as a function of energy with respect to the Coulomb barrier for  $^{6,7}$Li~+~$^{198}$Pt, $^{12}$C~+~$^{198}$Pt, $^{16}$O~+~$^{208}$Pb systems.  The calculated values correspond to the standard coupled-channels calculations using the code {\sc CCFULL} (b) Ratio of the slopes of L(E) and L$_{cs}$(E)  calculated at their crossing point,  as a function of the charge product of the reactants (Z$_1$.Z$_2$), for data from the present measurement and literature~\cite{review2,jia02,jia04,maha07,jia05,stef10,jia10,jiang14,araPRL,jia06}. Filled and open circles represent the data and extrapolated values obtained from the fit to L(E), respectively. The dashed line is obtained by fitting the data (filled circles) to an exponential function. } 
\label{fig:fig3}
\vskip -5mm
\end{figure}

    To further investigate the evolution of the fusion hindrance for different entrance channels, the ratio of the slopes of the logarithmic derivatives, $R$ = (dL(E)/dE)/(dL$_{cs}$(E)/dE) at the cross-over point between the L(E) and L$_{cs}$(E)~\cite{jia06,Jia07}, as a function of    Z$_1$.Z$_2$ is plotted in Fig.~3(b). The data used are from ~\cite{review2,jia02,jia04,maha07,jia05,stef10,jia10,jiang14,araPRL, jia06} and the present measurements. The ratio $R$ is a measure  of the fusion hindrance.  If the ratio approaches unity, the logarithmic slope of the data approaches the value for a constant S-factor and the sub-barrier hindrance can be considered to be absent while  larger values of $R$ indicate that the fusion cross section drops more rapidly implying a large hindrance~\cite{jia06,Jia07}.  The quantity Z$_1$.Z$_2$ is related to the strength of the coupling between the relative motion and the internal degrees of freedom. That is, when the nuclear coupling strength is estimated at the barrier position, it is proportional to Z$_1$.Z$_2$ in the linear coupling approximation, where the nuclear coupling form factor is proportional to $dV_N/dr$ (notice that $dV_N/dr=-dV_C/dr$ at the barrier position) \cite{review}. A strong correlation can be seen between $R$ and Z$_1$.Z$_2$ for different target projectile combinations (Fig.~3(b)).  Such a correlation was shown previously in Ref.~\cite{jia06,Jia07}. It was pointed out that the weaker hindrance with decreasing charge product implies that reactions of astrophysical interest are unlikely to be hindered. If the fusion hindrance is due to the damping of the coupling to collective motion, as the adiabatic model suggests, then  the effect of hindrance is expected to be small for lower values of Z$_1$.Z$_2$. 

 The trend of the fusion hindrance seen in Fig.~3(b), for reactions with light projectiles,  is expected to have an impact on  the synthesis of light elements  in astrophysical environment.  For energies relevant to astrophysical interest, the reaction rates are obtained from the  extrapolated  S-factor.  In Ref.~\cite{jia07} a method was proposed to extrapolate  S-factors for lighter systems, using the hindrance effect  observed  in  heavier systems. The results from this method  show that the  presence of the fusion hindrance can change the abundance of many isotopes in massive late-type stars, reduce reaction rates for carbon and oxygen fusion reactions (eg. $^{12}$C~+~$^{12}$C, $^{12}$C~+~$^{16}$O, and $^{16}$O~+~$^{16}$O) on stellar burning and nucleosynthesis~\cite{gasq07}. Based  on the correlation observed in Ref.~\cite{jia06} and shown in Fig. 3(b), including the new measurements, the fusion hindrance for such light  systems is expected to be weaker  than  those for heavy systems at energies corresponding to the peak of the S-factor.  At energies just below the S-factor peak, the sudden and adiabatic model calculations show different behaviors  for the heavier systems.  The calculations of the sudden model fall off steeply below the peak of the S-factor implying a strong hindrance. In contrast   a much weaker energy dependence of S(E) is expected from the adiabatic model~\cite{review,review2}.  At present, calculations from  both the sudden and the adiabatic models are not available at energies of astrophysical interest, close to the Gamow peak. It will be interesting to extend these calculations to the relevant energies. The reliability of such theoretical prediction  can only be confirmed when cross-sections for light ion fusion reactions, from challenging measurements at low energies will become available.
 
\section{Conclusion}
 In summary, the occurrence of fusion hindrance is clearly observed in case of $^{12}$C~+~$^{198}$Pt. The  adiabatic model calculation indicates a weak effect of the damping  for the present system  as compared to that obtained for systems with heavier projectiles.  On the other hand fusion hindrance has not been observed in case of $^{7}$Li~+~$^{198}$Pt, within the measured energy range. The corresponding  threshold energy estimated from the present measurement is found to be lower than the predicted values~\cite{hag07,jia09}.  The fusion hindrance   at  energies  deep  below   the barrier becomes progressively  significant in going from the    light    ($^{6,7}$Li) to  heavier ($^{12}$C, $^{16}$O) projectiles.   A strong correlation has been obtained  between  the degree of hindrance and the charge product over a wide range of  target-projectile combinations.  The observed trend reveals a weaker influence of hindrance on fusion involving lighter nuclei. This result together with a nearly flat energy dependence of S(E) in the adiabatic model at very low energies,   implies  that the effect of fusion hindrance will be  less substantial on astrophysical reaction rates for the production of light elements in stellar environments. New measurements of fusion cross-sections involving low Z elements including those  of astrophysical relevance, and   extension of existing theoretical models that explain fusion hindrance to the  energies close to the Gamow peak would be of interest.

\section{Acknowledgments}
 We are thankful to the accelerator staff for a smooth operation. AN acknowledges the support through the LIA France-India agreement.


\begin{thebibliography}{999}
\bibitem{review} K. Hagino and N. Takigawa, Prog. Theor. Phys. 128 (2012) 1001. 
\bibitem{review2} B.B. Back,  H. Esbensen, C.L. Jiang and K.E. Rehm, Rev. Mod. Phys.  86  (2014) 317, and references therein.
\bibitem{dissi-prl} A.O. Caldeira and A.J. Leggett, Phys. Rev. Lett.  46 (1981) 211; Ann. of Phys. 149
(1983) 374.
\bibitem{jia02} C.L. Jiang {\it et al.}, Phys. Rev. Lett.  89 (2002) 052701.
\bibitem{jia04} C.L. Jiang {\it et al.}, Phys. Rev. Lett. 93   (2004) 012701.
\bibitem{esb03} S. Mi\c{s}icu and H. Esbensen, Phys. Rev. Lett.  96 (2006) 112701; {\it ibid.}  Phys. Rev. C 75 (2007)  034606.
\bibitem{washiyama} K. Washiyama and D. Lacroix, Phys. Rev. C  78 (2008) 024610; A.S. Umar and V.E. Oberacker, Phys. Rev. C  74 (2006)  021601(R).
\bibitem{hag07} T. Ichikawa, K. Hagino and A. Iwamoto, Phys. Rev. C  75 (2007) 057603; {\it ibid.} Phys. Rev. C  75 (2007)   064612.
\bibitem{ichikawa2009} T. Ichikawa, K. Hagino and A. Iwamoto, Phys. Rev. Lett.   103  (2009) 202701;   {\it ibid.}  
Prog. theor. Phys. suppl.  196  (2012) 269. 
\bibitem{ichikawa_sub} Takatoshi Ichikawa,  Phys. Rev. C 92 (2015) 064604.
\bibitem{ichikawa2013} Takatoshi Ichikawa and Kenichi Matsuyanagi,  Phys. Rev. C   88 (2013) 011602(R).
\bibitem{ichikawa2015} Takatoshi Ichikawa and Kenichi Matsuyanagi, Phys. Rev. C 92 (2015) 021602(R). 

\bibitem{Daiz08} A. Diaz-Torres {\it et al.},   Phys. Rev. C   78 (2008) 064604; Phys. Rev. C  81 (2010) 041603(R).
\bibitem{maha07} M. Dasgupta {\it et al.}, Phys. Rev. Lett.  99 (2007) 192701.

\bibitem{jia05} C.L. Jiang {\it et al.}, Phys. Rev.C  71 (2005) 044613.
\bibitem{stef10} A. M. Stefanini  {\it et al.}  Phys. Rev. C   82 (2010) 014614.
\bibitem{jia10} C.L. Jiang {\it et al.}, Phys. Rev. C  82 (2010) 041601.
\bibitem{mont2010} G. Montagnoli {\it et al.},  Phys. Rev. C 85 (2012) 024607; {\it ibid.} 82 (2010) 064609.
\bibitem{mont2013}  G. Montagnoli {\it et al.},  Phys.Rev. C 87 (2013) 014611.
\bibitem{stef2014} A. M. Stefanini  {\it et al.}, Phys. Lett. B 728 (2014) 639. 
\bibitem{jiang14} C.L. Jiang {\it et al.}, Phys. Rev. Lett.  113 (2014) 022701.

\bibitem{araPRL} A. Shrivastava {\it et al.}, Phys. Rev. Lett.  103  (2009) 232702.
\bibitem{ant09} A. Lemasson {\it et al.},  Nucl. Instr. Meth. A  598  (2009) 445.

\bibitem{gav80} A. Gavron, Phys. Rev. C   21  (1980) 230.
\bibitem{araPRC} A. Shrivastava {\it et al.} Phys. Rev. Lett.  82 (1999) 699, {\it ibid.} Phys. Rev. C   63  (2001) 054602.
\bibitem{yao06} W.-M. Yao {\it et al.}, J. Phys. G  33 (2006) 1.
\bibitem{araPLB} A. Shrivastava {\it et al.}, Phys. Letts.  B  718  (2013) 931.
\bibitem{jia06} C.L. Jiang {\it et al.}, Phys. Rev. C   73 (2006)  014613.  
\bibitem{hagino} K. Hagino {\it et al.},  Comp. Phys. Comm.    123 (1999) 143.
\bibitem{jia09} C.L. Jiang {\it et al.}, Phys. Rev. C  79 (2009)  044601.
\bibitem{yas83} M. Yasue {\it et al.}, Nucl. Phys. A  394  (1983) 29.
\bibitem{Ei12} Ei Shwe Zin Thein, N. W. Lwin, and K. Hagino, Phys. Rev. C    85  (2012) 057602.
{\bibitem{Jia07} C.L. Jiang {\it et al.}, Phys. Rev. C  75 (2007) 057604.}
\bibitem{bala} A.B. Balantekin and P.E. Reimer,  Phys. Rev. C  33 (1986)  379; C.V.K. Baba, Nucl. Phys. A  553  (1993) 719c.
\bibitem{morton} C.R. Morton {\it et al.}, Phys. Rev. C  60 (1999) 044608.
\bibitem{Gomes} L.F. Canto, P.R.S. Gomes, R. Donangelo, J. Lubian, M.S. Hussein, Phys. Rep. 596 (2015) 1.
\bibitem{yang} X.P. Yang, G. L. Zhang and H. Q. Zhang {\it et al.}, Phys. Rev. C 87 (2013) 014603.
\bibitem{luong} D. H. Luong {\it et al.}, Phys. Lett. B 695  (2011) 105.

\bibitem{ara06} A. Shrivastava {\it et al.}, Phys. Lett. B 633  (2006) 433.

\bibitem{jia07} C.L. Jiang {\it et al.}, Phys. Rev. C  75 (2007) 015803.
\bibitem{gasq07} L.R. Gasques {\it et al.}, Phys. Rev. C 76 (2007) 035802.



\end{thebibliography}
\end{document}